\documentclass[prl,aps,twocolumn,showpacs,superscriptaddress,floatfix]{revtex4}
\usepackage{graphicx}
\usepackage{bm}

\newcommand{\beq}{\begin{equation}}
\newcommand{\eeq}{\end{equation}}
\newcommand{\bea}{\begin{eqnarray}}
\newcommand{\eea}{\end{eqnarray}}

\newcommand{\bq}{{\mathbf q}}

\def\tit#1#2#3#4#5{{#1}{\bf #2}, #3 (#4)}

\begin{document}

\title{Valence Bond Solids and
Their Quantum Melting
in Hard-Core Bosons on the Kagome Lattice}

\author{S. V. Isakov}
\affiliation{Department of Physics, University of Toronto, Toronto,
Ontario M5S 1A7, Canada}
\author{S. Wessel}
\affiliation{Institut f\"ur Theoretische Physik III, Universit\"at
  Stuttgart, 70550 Stuttgart, Germany}
\author{R. G. Melko}
\affiliation{Materials Science and Technology Division, Oak Ridge
National Laboratory, Oak Ridge, TN 37831, USA}
\author{K. Sengupta}
\affiliation{ TCMP division, Saha Institute of Nuclear Physics, 1/AF
Bidhannagar, Kolkata-700064, India}
\author{Yong Baek Kim}
\affiliation{Department of Physics, University of Toronto, Toronto,
Ontario M5S 1A7, Canada}

\date{\today}

\begin{abstract}

Using large scale quantum Monte Carlo simulations and dual vortex theory
we analyze the ground state phase diagram of hard-core bosons on the kagome
lattice with nearest neighbor repulsion. In contrast to the case of
a triangular lattice, no supersolid emerges for strong interactions. While
a uniform superfluid prevails at half-filling, two novel solid phases emerge
at densities $\rho=1/3$ and $\rho=2/3$. These solids exhibit an only partial
ordering of the bosonic density, allowing for local resonances on a subset
of hexagons of the kagome lattice. We provide evidence for
a weakly first-order phase transition at the quantum melting point 
between these solid phases and the superfluid.
\end{abstract}

\pacs{05.30.Jp, 75.10.Jm, 75.40.Mg, 71.27.+a}

\maketitle

Current interest in microscopic models of frustrated quantum systems
stems largely from the search for exotic quantum phases and spin
liquid states. In general,  geometric frustration tends to destabilize
quasi-classical order, possibly allowing for nontrivial quantum states
and novel critical phenomena to emerge in such systems. One
intriguing approach addresses classically frustrated (Ising) models
perturbed by quantum (off-diagonal) interactions \cite{moessner1}.
The behavior of classically disordered, degenerate ground state
manifolds upon application of a U(1) symmetric perturbation (e.g.
ferromagnetic exchange) is of special interest as experimental
advances in
the construction and control of atomic gases in optical lattices have
opened up the possibility of designing such Hamiltonians for ultracold
bosons. In particular, it has recently been shown how
an optical kagome lattice can be constructed using a triple laser
beam design \cite{Santos}, which could permit access to parameter regions
of interest in the search for exotic quantum phenomena.

\begin{figure}[t]
\includegraphics[width=2.54in]{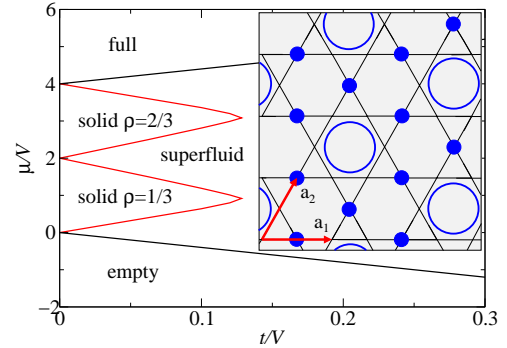}
\caption{ (color online).
Ground state phase diagram of hard-core bosons on the kagome
lattice (inset).  The primitive vectors ${\bf a}_1$ and ${\bf a}_2$ are
constrained on a (periodic) torus spanned by
${\bf L}_1 = n_1 \times {\bf a}_1$ and ${\bf L}_2 = n_2 \times {\bf a}_2$
(where $a_1=a_2=2$). The circles illustrate
the subset of hexagons with a resonating boson occupation of three bosons per
hexagon in the $\rho=2/3$ solid. The remaining bosons localize to form a
solid backbone on the sites that do not belong to any of these hexagons.
}
\label{kaglatt}
\end{figure}

In this paper we consider a model of bosons on the kagome lattice in the
strongly interacting regime, corresponding to the hard-core limit of the
Bose-Hubbard Hamiltonian discussed in Ref.~\cite{Santos},
\beq
  H_{b}=-t \sum_{\langle i,j \rangle}
    \left( b^{\dag}_i b_j + \text{h.c.} \right)
    + V \sum_{\langle i,j \rangle} n_i n_j - \mu \sum_{i} n_i,
\eeq where $b^\dagger_i$ ($b_i$) creates (destroys) a particle on
site $i$, $t>0$ denotes the nearest-neighbor hopping, $V>0$ is the
nearest-neighbor repulsion, and $\mu$ is the chemical potential.
This model can also be mapped onto the spin-$1/2$ XXZ model~\cite{mapping}, 
allowing for an interpretation of
our results in terms of both bosons
and quantum spins.  We report results on the ground state phase diagram
obtained from a combined analysis of large scale quantum Monte Carlo (QMC)
simulations using the stochastic series expansion
technique~\cite{sse1,more_details} and  phenomenological dual
vortex theory (DVT) \cite{dual,dualkag}.  We find that, in contrast to 
previous theoretical expectations, a uniform superfluid persists at
half-filling for all values of $V/t$.  In addition, for fillings $\rho=1/3$
and $2/3$ we find evidence for valence bond solid (VBS) phases where bosons
are delocalized around a subset of hexagons (see Fig.~\ref{kaglatt}). We find 
that the quantum melting of both VBS phases into the superfluid 
occurs at weakly first-order quantum phase transitions.

Past work on the ground state phase diagram of this model
has been controversial and intriguing:  Spin-wave calculations suggest that
a supersolid state may emerge around
half-filling ($\rho=1/2$) at  $\mu=2V$ for $t/V<0.5$~\cite{Murthy97}.
However, these results are not conclusive since strong quantum fluctuations
may destroy the long-range order assumed within mean-field
theory~\cite{Murthy97}. More recently, consideration of the large classical
degeneracy~\cite{ClassIsing} at $t=0$ has led to the proposal of several
exotic Mott-insulating states (e.g.~VBSs or disordered quantum liquids) at
half-filling~\cite{NikSen}.  Very little work has been done to elucidate the
nature of the phase diagram away from half-filling.

\begin{figure}[floatfix]
{
\includegraphics[width=1.4in]{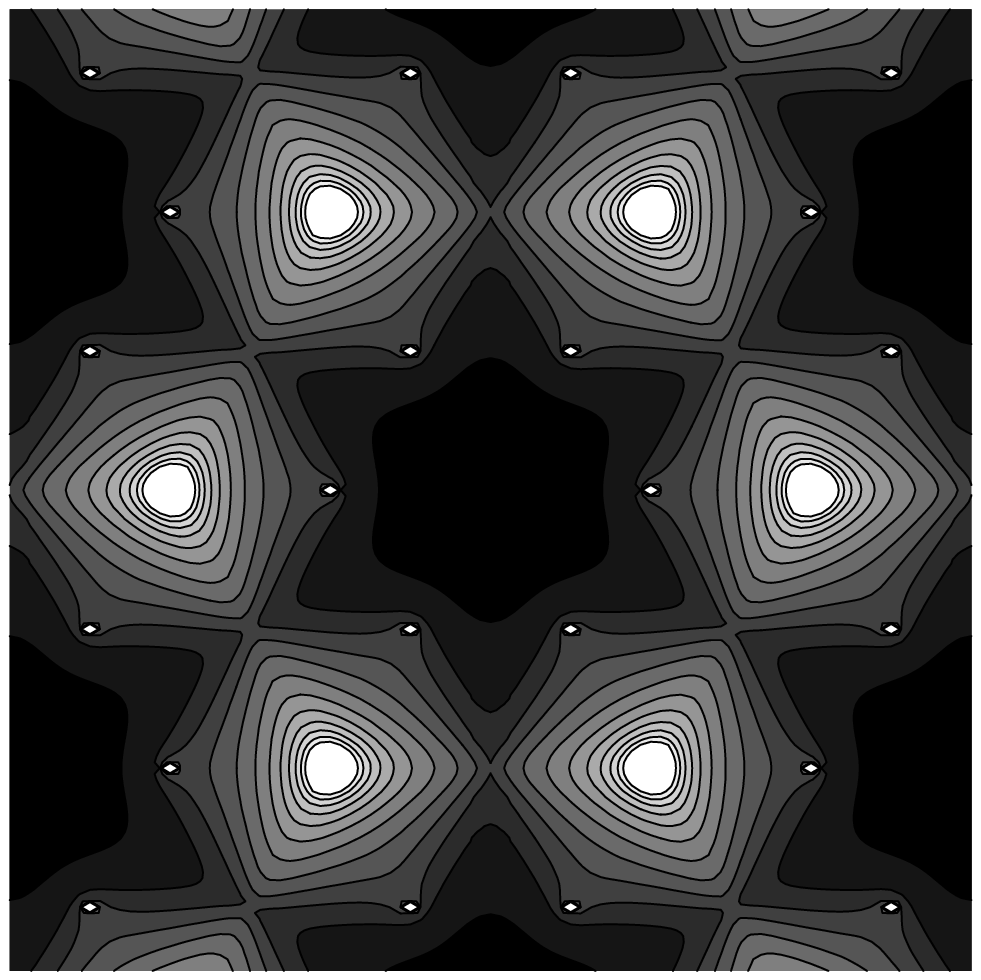}
\includegraphics[width=1.4in]{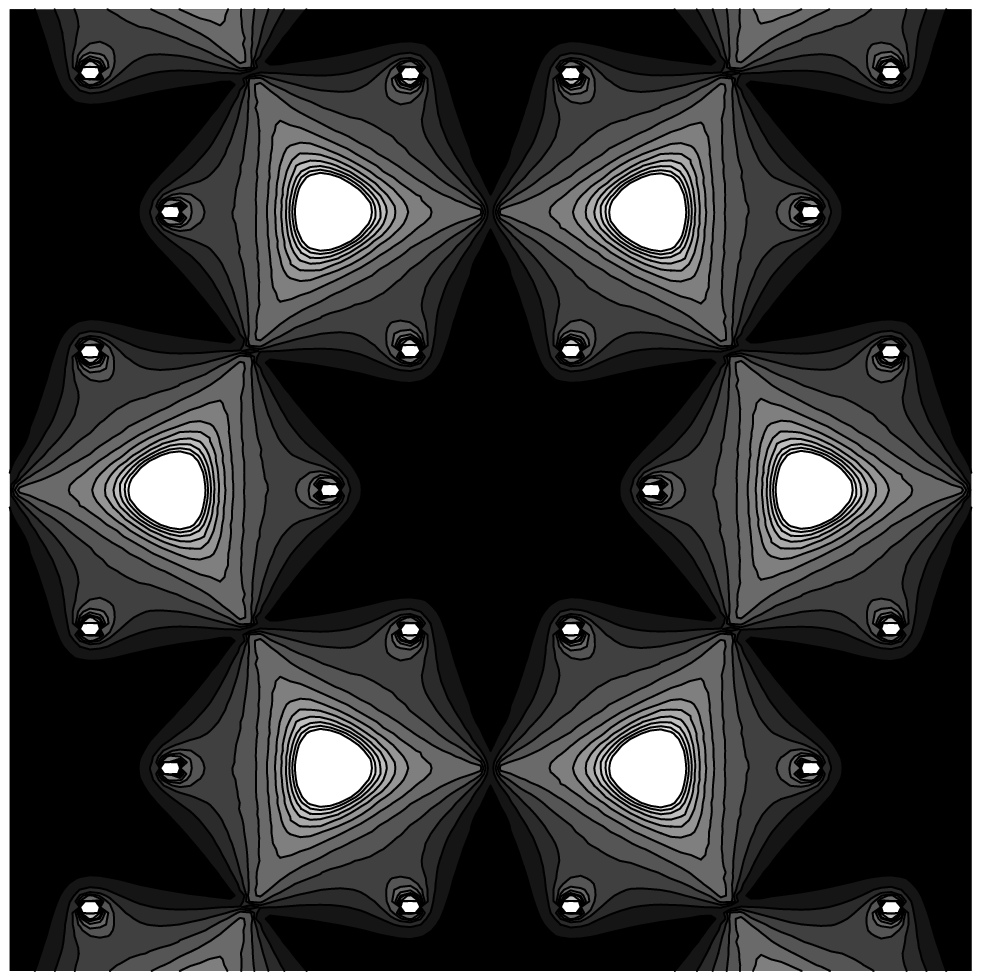}
\caption{Contour plots of the structure factor (left panel) and the
static susceptibility (right panel) in the VBS close to the phase
boundary for $\mu/V=11/12$, $t/V=1/8$, and $T=t/24$. The BZ of the
underlying triangular lattice is hexagonal (see Fig.~\ref{kaglatt}), with
corners at ${\mathbf Q}=(2\pi/3,0)$ and symmetry related momenta.  In both
panels the  axes range from  $-2\pi$ to $2\pi$, and ferromagnetic peaks at
reciprocal lattice vectors have been suppressed.
\label{fig:solid}}
}
\end{figure}

Using QMC simulations, we have obtained the phase diagram of $H_b$,
illustrated in Fig.~\ref{kaglatt}. The lattice is empty for $\mu\leq -4t$
and completely filled for $\mu\geq 4(t+V)$. For large values of  $t/V$, the
bosons are superfluid, with a finite value of the superfluid density
$\rho_s$, which we measure through winding number fluctuations
$W_{a_{1,2}}$~\cite{windingnumber} in each of the lattice directions as
$\rho_s=(\langle W_{a_1}^2\rangle + \langle W_{a_2}^2 \rangle)/({2 \beta t})$,
where $\beta$ is the inverse temperature.
In agreement with mean-field theory, we find that two solid phases
with rational fillings $\rho=1/3$ and $2/3$ emerge at smaller $t/V$. 
Both are characterized by finite values (in the thermodynamic limit) of the
density structure factor per site, $S({\mathbf q})/N=\langle
\rho_{{\mathbf q}\tau} \rho^{\dagger}_{{\mathbf q}\tau} \rangle$ and
the static susceptibility per site, $\chi({\mathbf q})/N=\langle
\int d\tau\rho_{{\mathbf q}\tau} \rho^{\dagger}_{{\mathbf q}0}
\rangle$, where $\rho_{{\mathbf
q}\tau}=(1/N)\sum_i\rho_{i\tau}\exp(i{\mathbf q}{\mathbf r_i})$ and
$\rho_{i\tau}$ is the boson density at site $i$ and imaginary time
$\tau$, at wave vectors ${\mathbf q}={\mathbf Q} \equiv (2\pi/3,0)$. This can
be seen in Fig.~\ref{fig:solid}, which shows peaks at the corners of the 
Brillouin zone (BZ) in the solid phase that are absent in the
superfluid. 

\begin{figure}[floatfix]
{
\centerline{\includegraphics[width=2.8in]{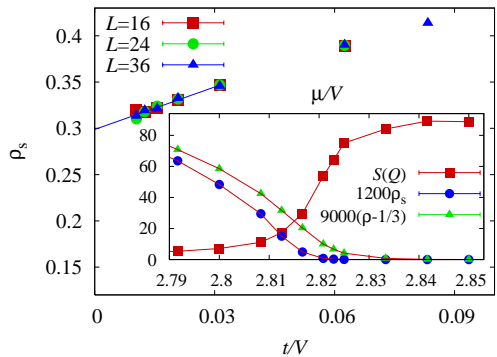}}
\caption{ (color online).
The superfluid density $\rho_s$ as a function of $t/V$ at half-filling for
$\mu/V=2$ and T=t/5. The lattice linear dimension is denoted $L$
(see Fig.~\ref{kaglatt}), so that the number of sites is
$N = L \times L \times 3$. Inset: The static structure factor
$S({\mathbf Q})$, superfluid density $\rho_s$, and boson density $\rho$ as
a function of $\mu/V$ at $t/V=1/10$, for $L=24$ and $T=t/24$.
\label{fig:rhos:f:1:2}}
}
\end{figure}

Decreasing $t/V$ at half-filling ($\mu/V=2$), the value of $\rho_s$
in the limit $V\rightarrow \infty$ ($t=1$) takes on about 54\%  of its
value at the XY point, $\rho_s(V=0)\approx 0.55$
(see Fig.~\ref{fig:rhos:f:1:2}).
This is in stark contrast to the triangular lattice case, where
$\rho_S(V\rightarrow \infty)$ approaches  only about 4\% of its value at
the XY point~\cite{sstri2}.  We do not observe
evidence of long-range order in the density structure factor, thus
eliminating the possibility of a supersolid phase as found recently on the
triangular lattice at small values of
$t/V$~\cite{sstri1, sstri2, sstri4}.  
In addition, we do not observe any Bragg peaks in the bond-bond structure
factor (defined below), precluding the existence of any VBS order.
The persistence of the superfluid phase can be understood from a duality
analysis of the boson problem in terms of vortices on the dual dice lattice,
which interact with a dual magnetic flux of $2\pi p/q = \pi$ (for boson
filling $\rho = p/q =1/2$) \cite{dual,dualkag}. It has been shown that for
the dice lattice at $\rho=1/2$ the vortices undergo dynamic localization
due to an Aharanov-Bohm caging effect \cite{vidal1} that suppresses
the condensation of vortices and leads to a persistence of superfluidity
\cite{dualkag}.

\begin{figure}[floatfix]
{
\centerline{\includegraphics[width=2.8in]{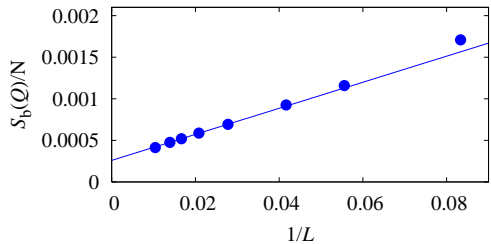}}
\caption{ (color online).
Finite-size scaling of the bond structure factor $S_{\text b}({\mathbf Q})$
in the VBS phase for $\mu/V=11/12$, $t/V=1/8$, and $T=t/12$. Error bars are
smaller than the symbol sizes, and the line shows a linear extrapolation to
the thermodynamic limit.
\label{fig:bsf}}
}
\end{figure}

\begin{figure}[floatfix]
{
\centerline{\includegraphics[width=2.8in]{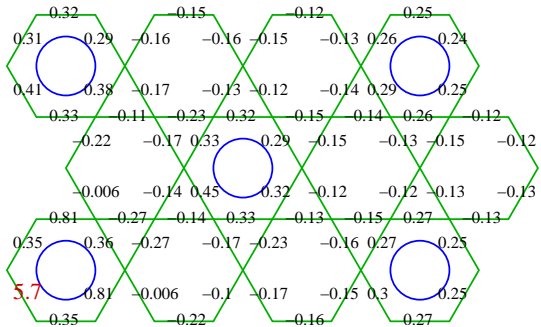}}
\caption{ (color online).
The correlation function $C_{\text b}({\mathbf r}_1-{\mathbf r}_\delta)$ between
the bond indicated by large text (red online) and the other bonds for $L=24$,
$\mu/V=11/12$, $t/V=1/8$, and $T=t/12$. 
\label{fig:bchi:rs}}
}
\end{figure}

At boson fillings $\rho=1/3$ and $2/3$, the model in the large
$V/t$ limit can be mapped to a quantum dimer model on the hexagonal
lattice, where the occupied (unoccupied) sites at $\rho=1/3$
($\rho=2/3$) correspond to the presence of dimers.  In the classical limit
($V/t=\infty$) all of the dimer coverings are degenerate. When $V/t$ becomes
large but finite, the degeneracy is partially lifted, and the remaining
dimer kinetic energy promotes the so-called plaquette order
\cite{MSCprb2001,cabra}.
In terms of bosons,  for example at $\rho=2/3$, every third site on the
kagome lattice forms a solid backbone of occupied sites, whereas
the remaining bosons locally resonate around every third
hexagon, as shown in Fig.~\ref{kaglatt}.
DVT at $\rho=1/3$ and $2/3$ predicts several possible
valence bond order phases on the kagome lattice depending on
the parameters of the effective field theory of the dual vortices.
A set of mean-field states obtained by DVT are found to have
the same spatial symmetry as the state described above.
We expect that the kinetic energy gain would prefer
the phase with the resonating hexagons.
Our QMC data on $S(\bq)$ and $\chi(\bq)$
is consistent with these expectations; in particular, Bragg peaks are
observed at the correct positions, as shown in Fig.~2.
We find further QMC evidence for such resonances by measurements of the
hexagon occupation. Denoting by $P_n$ the measured fraction
of hexagons with $n$ bosons, we obtain
$P_{0,1,2}<10^{-3}$,
$P_3=0.33(1)$,
$P_4=0.38(2)$,
$P_5=0.225(4)$,
$P_6=0.052(4)$ at various points inside the  $\rho=2/3$ solid, which compare
well to the expected values $P^e_n$ of
$P^e_{0,1,2}=0$,
$P^e_3=5/12$,
$P^e_{4,5}=1/4$,
$P^e_6=1/12$ for the resonating state.
Finally, we find sharp peaks in the bond structure factor and corresponding
susceptibility at the corners of the kagome BZ. Fig.~\ref{fig:bsf} shows 
the finite size scaling properties of the structure factor
$S_{\text b}({\mathbf q})/N=\langle
B_{{\mathbf q}\tau} B^{\dagger}_{{\mathbf q}\tau} \rangle$ at ${\mathbf
q}={\mathbf Q}$, where
$B_{{\mathbf q}\tau}=(1/N)\sum_\alpha B_{\alpha \tau}
\exp(i{\mathbf q}{\mathbf r_\alpha})$ summed over the bond index $\alpha$
connecting spins $i$ and $j$, and
$B_{\alpha(i,j), \tau}=t(b^{\dag}_i b_j + b_i b^{\dag}_j)_{\tau}$ is
the off-diagonal bond operator at imaginary time $\tau$. The non-zero
intercept of the extrapolation in  Fig.~\ref{fig:bsf} exhibits long
range order in the bond-bond correlations~\cite{footnote}. 
The real-space correlations,
$C_{\text b}({\mathbf r}_\gamma-{\mathbf r}_\delta)=\langle
(\frac{1}{\beta}\int B_{\gamma\tau}d\tau-B_0)
(\frac{1}{\beta}\int B_{\delta\tau}d\tau-B_0) \rangle$,
where $B_0$ denotes the background bond strength, confirm the expected
preponderance of resonating bonds on hexagons as illustrated in
Fig.~\ref{fig:bchi:rs}.

As shown in Fig.~\ref{fig:solid}, 
``bow-tie'' features appear in $S(\bq)$ and
$\chi(\bq)$. At low temperatures and close to
${\mathbf q}_0=(0,2\pi/\sqrt{3})$ and equivalent positions, these can be
fitted to
$S(\bq_0+\bq) \sim (q_{\parallel}^2+\kappa^2)/\sqrt{q^2+\kappa^2},$
$\chi(\bq_0+\bq) \sim (q_{\parallel}^2+\kappa^2)/(q^2+\kappa^2),$
where $q_{\parallel}$ is a component of
$\bq$ along the directions of bow-ties, and $\kappa$ is interpreted
as an inverse correlation length. The bow-ties are present in
the solid phase and also in the superfluid phase near the transition, albeit
with larger values of $\kappa$. Implications of this bow-tie structure will
be discussed elsewhere~\cite{more_details}.

In contrast to the case of a triangular lattice~\cite{sstri1}, we do not
observe any obvious discontinuities that would indicate a first-order phase
transition from the solid phases into the superfluid. An example is illustrated
in Fig.~\ref{fig:rhos:f:1:2}, for $\rho_s$, $S({\mathbf Q})$
and $\rho$ along a cut at fixed $t/V$ from the superfluid into the $\rho=1/3$
solid. Landau-Ginzburg-Wilson (LGW) theory forbids a generic continuous
transition between a solid and a superfluid, as different symmetries of $H_b$
are broken in both phases. Possible explanations of our observation would
thus include (i) a {\em weakly} first-order superfluid-solid transition, at
which the correlation length stays finite but exceeds the linear
size of the QMC cells; (ii) a narrow intermediate supersolid regime; or (iii)
a non-LGW continuous transition, such as the recently proposed deconfined
quantum criticality scenarios ~\cite{deconf1,deconf2}. 
To discern between these possibilities
we obtained detailed data, and performed a finite-size scaling (FSS) study and
analysis of energy histograms over the melting transition region.

\begin{figure}[floatfix]
{
\centerline{\includegraphics[width=2.8in]{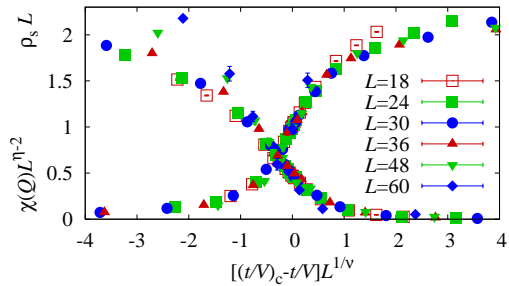}}
\caption{ (color online).
Data collapse of the superfluid density $\rho_s$ and static susceptibility
$\chi({\mathbf Q})$ at $\mu/V=11/12$ for $\beta/L=1/t$.
}
\label{fig:collapse}
}
\end{figure}

We look first at the FSS behavior.
For a two-dimensional system, the following FSS relations
apply for a {\it continuous} solid to superfluid quantum phase transition. The
superfluid density scales as
$
  \rho_s=L^{-z} F_{\rho_s}(L^{1/\nu}(K_c-K), \beta/L^z),
$
where $L$ denotes the linear system size, $z$ the dynamical critical
exponent, $\nu$ the correlation length exponent, $K_c-K$ the
distance to the critical point in terms of the control parameter $K$, such
as $t/V$ or $\mu/V$, and $F_{\rho_s}$ the corresponding scaling function.
Similarly,
$
  S({\mathbf Q})=L^{2-z-\eta} F_{S}(L^{1/\nu}(K_c-K), \beta/L^z) 
$ and
$
  \chi({\mathbf Q})=L^{2-\eta} F_{\chi}(L^{1/\nu}(K_c-K), \beta/L^z),
$
where $\eta$ is the anomalous exponent. 
From these scaling relations, given appropriate values
for $(t/V)_c$ and the critical exponents, QMC data for different system
sizes should follow universal curves $F_{\rho_s}(\cdot,A)$ and
$F_{\chi}(\cdot,A)$, if the transition is continuous.

QMC data was obtained over the melting transition, focusing on a fixed
value of $\mu/V=11/12$, which is close to the largest extend of the
$\rho=1/3$ solid.  
As discussed below, the scaling of the QMC data appears consistent with $z=1$,
so that most simulations were performed for fixed aspect ratios $A=\beta/L$.
As shown in Fig.~\ref{fig:collapse},
the data for $\rho_s L$ appears to collapse very well for
$(t/V)_c=0.12821(2)$, $z=1$, and $\nu=0.43$. The data for $\chi({\mathbf Q})$
also collapses well for $(t/V)_c=0.12827(4)$, $\nu=0.45$, and $\eta=-0.50(15)$.
The two independent estimates for $(t/V)_c$
are sufficiently close in value 
to confirm a direct transition and make scenario
(ii) seem unlikely.  This is also consistent with the discovery of similar
$\nu$ values on both sides of the transition.  Based on the value of
$\eta=-0.50(15)$ we also find the
scaling of $S({\mathbf Q})$ to be consistent with a dynamical
exponent $z=1.0(2)$. 

\begin{figure}[floatfix]
{
\centerline{\includegraphics[width=2.8in]{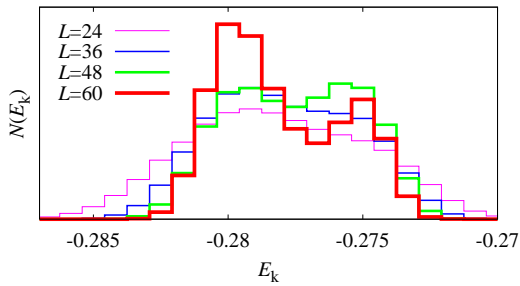}}
\caption{ (color online).
Distribution (arbitrary units) of the kinetic energy close to the
transition point for various system sizes, $\mu/V=11/12$, $t/V=1/7.796$,
and $T=t/60$.  As the system size increases, the double-peaked
features become more pronounced.
}
\label{fig:histogram}
}
\end{figure}

Even though the FSS analysis appears consistent with a continuous quantum
critical point in most respects,  a negative value of $\eta$ is unusual,
and one may be inclined to interpret this as an indication of very {\it weak}
first-order behavior.
We explore this possibility in further detail by studying histograms of physical
quantities in the 
transition region.  In Fig.~\ref{fig:histogram}, one can see double-peaked
structures 
developing in histograms of the kinetic energy over the melting transition
for sufficiently large system sizes.  We believe that this additional
observation provides
cogent evidence for scenario (i), i.e.~the melting transition studied here is
indeed very weakly first-order.  

In conclusion, we have studied hard-core bosons on the kagome lattice with
nearest-neighbor repulsion. At half-filling the
model remains in a uniform superfluid phase for all values of $V/t$, and no
supersolid emerges. For the solid phases at
filling fractions $1/3$ and $2/3$, we find evidence for an exotic insulator
with partially delocalized bosons occurring on a six-site hexagonal structure
on the kagome lattice.
Although the superfluid-solid melting transition appears naively to have the
scaling properties
of an unusual continuous quantum critical point, we find clear indication
of weak first-order behavior from double-peaked histograms of the kinetic
energy. The apparent weakness of this transition is in stark contrast 
to examples of strong first-order quantum melting transitions in related
models~\cite{sstri1}.
An understanding of this contrasting behavior may have critical importance in
the search for
unconventional quantum criticality in this class of models, and clearly deserves
further study in the future.

This work was supported by the NSERC of Canada, CRC, CIAR (SVI and YBK) as well
as NIC J\"ulich and HLRS Stuttgart (SW). We  thank R. Moessner, A. Paramekanti,
A. Vishwanath, T. Senthil, A. Sandvik, S. Sachdev, A. Honecker, K. Damle,
N. Prokof'ev, and B. Svistunov for helpful discussions.

{\it Note added.}-- Recently, we became aware of a parallel work
\cite{Damle}, and would like to thank its authors for correspondence.


\begin{thebibliography}{99}
\bibitem{footnote}
The extrapolated value of order $10^{-4}$ appears quite small.  
However, the bond structure factor quantifies an
off-diagonal correlation function and order parameters constructed from such
quantities typically are reduced in magnitude (from unity) due
to quantum fluctuations. This value can be compared to the only other value
available in the literature \cite{sandvik}, where deep in the VBS phase the
bond structure factor extrapolates to $\sim 10^{-3}$.
\end{thebibliography}
\end{document}